\documentstyle[epsf,aps,floats]{revtex}
\begin{document}
\draft
\title{A nonlinear detection algorithm for periodic signals in gravitational wave detectors}
\author{Julien Sylvestre}
\address{LIGO Project, Massachusetts Institute of Technology,\\ NW17-161, 175 Albany St., Cambridge, MA 02139, USA.\\julien@ligo.mit.edu}

\date{\today}
\maketitle

\begin{abstract}
We present an algorithm for the detection of periodic sources of gravitational waves with interferometric detectors that is based on a special symmetry of the problem: the contributions to the phase modulation of the signal from the earth rotation are exactly equal and opposite at any two instants of time separated by half a sidereal day; the corresponding is true for the contributions from the earth orbital motion for half a sidereal year, assuming a circular orbit. 
The addition of phases through multiplications of the shifted time series gives a demodulated signal; specific attention is given to the reduction of noise mixing resulting from these multiplications.
We discuss the statistics of this algorithm for all-sky searches (which include a parameterization of the source spin-down), in particular its optimal sensitivity as a function of required computational power.
Two specific examples of all-sky searches (broad-band and narrow-band) are explored numerically, and their performances are compared with the stack-slide technique (P. R. Brady, T. Creighton, Phys. Rev. D {\bf 61}, 082001).
\end{abstract}


\section*{Introduction}
Kilometer-scale gravitational wave interferometric detectors will become operational by the end of 2001, allowing observations with unprecedented sensitivity at frequencies ranging from a few tens of hertz to approximately one kilohertz \cite{LIGO_VIRGO}.
Periodic sources form an interesting subclass of potentially detectable astrophysical objects, due to the possibility to improve the ``visibility'' of a weak periodic signal in noisy data by increasing the observation time.
Good candidates for such sources are spinning galactic neutron stars with a deformation misaligned with their rotation axis; optimistic estimates suggest that LIGO I could detect such objects with an observation time of the order of one year \cite{Thorne}.

Frequency modulations due to the Doppler shift induced by the detector motions, and intrinsic frequency variations due to the loss of angular momentum to gravitational waves, are two easily parameterizable examples of properties of a realistic signal that will dramatically increase the difficulty of the data analysis. In fact, for all but the simplest types of searches, the data analysis will be sub-optimal because of computational limitations.
More complications are likely to arise if the source is in a binary, due to the additional frequency modulation \cite{DV} and possibly larger or random intrinsic frequency variations expected if the source is accreting.

These considerations have driven serious efforts to define data analysis schemes that would maximize the sensitivity of a search (to be defined in section \ref{sensitivity}), given a maximal available computational power.
Matched filtering analysis, together with an in-depth discussion of signal properties, is presented in a series of papers \cite{JKS}; for all-sky searches, this technique would be computationally prohibitive.
The so-called coherent search, discussed in \cite{BCCS}, uses a resampling technique to demodulate the signal and to correct for intrinsic frequency variations: the detection is performed by taking the Fourier transform with respect to the resampling time, which amounts to a change of frame of reference from the detector to the solar system barycenter and to a ``stretching'' of time for frequency variations.
This technique is generally not practical for all-sky searches over long time periods (longer than a few days), but a refinement of it called the stack-slide technique \cite{BC} is considered as one of the best of all known techniques: the initial data are segmented into shorter intervals, and their power spectra are built using the coherent method.
For these shorter intervals, a smaller number of points in parameter space have to be used.
The spectra are added (incoherently) after having been shifted according to a finer gridding of parameter space. 
A related technique uses Hough transforms \cite{SP} to track the frequency evolution of peaks in the spectra from the shorter intervals, which is known in advance for some choice of the source parameters.
In general, it is possible to improve the sensitivity of these techniques by integrating them into more general hierarchical searches \cite{BC,SP}: a first search in parameter space is done at reduced sensitivity but low confidence level (so that spurious events are likely); the vicinity of candidate events is then scrutinized at full sensitivity to get higher confidence detections.

We have explored yet another approach that exploits a particular symmetry of the problem in an attempt to reduce its complexity.
Firstly, we note that the Doppler shift in the frequency of a given source due to the rotation of the earth, independently of the source position, is exactly equal and opposite at any two instants of time separated by half a sidereal day (we will only consider linear terms in the Doppler shift, an excellent approximation).
Secondly, in the approximation that the earth orbital motion around the sun is circular, the Doppler shift in frequency has the same property as above, but for a time separation of half a sidereal year.
We shall use this approximation for illustrative purposes; corrections for the small eccentricity of the earth orbit and for the influence of the moon and of other planets would have to be included in practice.
Signal phases at different times can be added by properly multiplying the (analytic extensions of the) signal time series, and therefore a signal free of frequency modulations may in principle be built.
We will give a more precise formulation of this idea in section \ref{algorithm}.

The multiplications of different stretches of the data will strongly affect the noise characteristics, principally by mixing noise at different frequencies; a useful implementation will have to bandpass filter the data to minimize this effect, and therefore a search over source parameters (frequency, frequency derivatives, source position) will be required.
An example of a possible implementation will be presented in section \ref{implementation}. 
The achievable sensitivity (to be defined in section \ref{sensitivity}) will be studied numerically as a function of computational power requirements in section \ref{numerical}, and comparisons with the stack-slide method mentioned above will be presented.

\section{Algorithm} \label{algorithm}
We will use the following model for the signal $s(t)$ at the detector:
\begin{eqnarray}
s(t)=h \cos[\phi(t)], \\
\phi(t)=2\pi f_0 [t + M_1(t) + M_2(t)], \label{eq:phase_fnc}\\
M_1(t)=\frac{A_1 {\rm d}}{2\pi} \cos\left(\frac{2\pi t}{{\rm d}} + \alpha \right), \\
M_2(t)=\frac{A_2 {\rm yr}}{2\pi} \cos\left(\frac{2\pi t}{{\rm yr}} + \psi \right), \\
A_1=\frac{2\pi R_\oplus}{{\rm d} \cdot c} \cos(\lambda) \cos(\delta), \label{eq:A1} \\
A_2=\frac{2\pi \rm{A.U.}}{{\rm yr} \cdot c} \cos(\delta+\epsilon), \label{eq:A2} 
\end{eqnarray}
where $\epsilon$ is the earth oblicity, $\lambda$ is the detector latitude, $c$ is the speed of light, $R_\oplus$ is the earth radius, A.U. is an astronomical unit, d is a sidereal day length, yr is a sidereal year length, $(\alpha,\delta)$ is the source angular position, $f_0$ is the signal frequency in the source rest frame (it may be a function of time), and $\psi$ is a fixed number that defines the origin of time.
For simplicity, since our focus is on the phase modulation of the signal, the explicit time dependence of $h$ on the detector response pattern will not be considered, although its averaged effect will be included in our definition of sensitivity (section \ref{sensitivity}).
If $f_0$ is independent of time, a good approximation to the signal half-bandwidth $\Delta f$ over a time interval of length $t$ (i.e., half the intrinsinc frequency change over $t$), is given by:
\begin{equation}
\Delta f(t) = f_0 [A_1\min(1,t/{\rm d}) + A_2\min(1,t/{\rm yr})]. \label{eq:signal_bandwidth}
\end{equation}
The numerical values of the terms multiplying the cosine terms in eq. \ref{eq:A1} and eq. \ref{eq:A2} are approximately $1.55 \cdot 10^{-6}$ and $9.94 \cdot 10^{-5}$, respectively.
Hence, the half-bandwidth of a 1 kHz source would be roughly 2 mHz over one day, and 0.1 Hz over a year.

The phase function $\phi(t)$ is sufficiently well-behaved that we can use the Hilbert transform of $s(t)$ to build its quadrature \cite{Hahn}, $\hat{s}(t)=h\sin[\phi(t)]$, and therefore construct the analytic signal $S(t)$:
\begin{equation}
S(t)=h e^{i\phi(t)}.
\end{equation}
Assuming that $f_0$ is independent of time, we have that
\begin{eqnarray}
S_4(t)=S(t)S(t+{\rm d}/2)S(t+{\rm yr}/2)S(t+{\rm d}/2+{\rm yr}/2) \nonumber \\
 = h^4 e^{8\pi i f_0 (t + {\rm d}/4 + {\rm yr}/4)},
\end{eqnarray}
that is, $S_4$ is monochromatic with frequency $4f_0$.
If $f_0$ is a function of time, however, $S_4$ as defined above will not be demodulated because of the coupling of $f_0$ with the modulations; its phase will be
\begin{eqnarray}
\frac{\phi_4(t)}{2\pi} = f_0(t)t + f_0(t_1)t_1 + f_0(t_2)t_2 + f_0(t_3)t_3 + \nonumber \\
M_1(t)[f_0(t)-f_0(t_1)] + M_1(t_2)[f_0(t_2)-f_0(t_3)] + \nonumber \\
M_2(t)[f_0(t)-f_0(t_2)] + M_2(t_1)[f_0(t_1)-f_0(t_3)], \label{eq:S4M}
\end{eqnarray}
where $t_1=t+{\rm d}/2$, $t_2=t+{\rm yr}/2$ and $t_3=t+{\rm d}/2+{\rm yr}/2$.
The signal so obtained still presents some residual phase modulation from the detector motions, but it results in a much smaller contribution $\Delta f_{\rm mod}$ to the signal bandwidth; the terms multiplying the modulation functions $M_1$ and $M_2$ are now of order of the intrinsic frequency {\it change} over half a day and half a year, respectively, rather than the source frequency itself.
We will show below that our algorithm best performances are obtained by dividing the total data from an observation of length $t_{\rm obs}$ into smaller segments of length $T$ shorter than one day.
Consequently, we impose the condition that the signal bandwidth produced by the residual modulations in eq.~\ref{eq:S4M} be smaller than the frequency resolution for a segment of length $T$, $\Delta f_{\rm mod} < 1/T$, assuming a linear frequency model: $f_0(t)=f_0 \cdot [1-t/\tau]$, where $\tau$ is the source spin-down characteristic time.
We find that this condition is:
\begin{equation}
\tau > f_0 T^2 (A_1 + A_2),
\end{equation}
i.e.,
\begin{equation}
\tau > \left(\frac{f_0}{1{\rm kHz}}\right) \left(\frac{T}{1 {\rm d}}\right)^2 \cdot 25 {\rm yr}.
\end{equation}
This estimate shows clearly that for most realistic sources, the demodulation by multiplication is performed correctly, so that we can assume that spin down enters the phase function only through the combination $f_0(t)t + f_0(t_1)t_1 + f_0(t_2)t_2 + f_0(t_3)t_3$.

Noise will be seriously affected by the multiplications of data segments at different times, through nonlinear mixing of signal with noise and of noise with noise at different frequencies.
The latter of these effects is so severe that we will have to minimize it as much as possible by bandpass filtering individually the data segments around the signal, or the technique will be of no utility.
An efficient filtering requires a knowledge of the frequency content of the signal as a function of time, and will therefore involve a search over source frequencies, sky positions and spin-down parameters; this will greatly increase the computational costs.
This search will be implemented by meshing this parameter space with a certain resolution, i.e. with a certain number of points $N_p$. 
Obviously, if a given signal has parameters that are not exactly coincident with any of these points, the sensitivity of our search to this signal will be reduced.
We will explore below how the density of the mesh couples to the achievable mean sensitivity, and therefore give an estimate of the usefulness of our technique.

\subsection{Implementation}    \label{implementation}
From the general guidelines established in the previous section, we are able to discuss an implementation of our algorithm. 
This implementation has parameters that allow an optimization of the method, subject to computational power constraints.

The first step will be to prepare for the analysis data obtained from an observation of length $t_{\rm obs}$.
The dataset is compressed to contain only information for frequencies of interest, i.e., frequencies between $f_{\rm min}$ and $f_{\rm max}$.
The analytic signal is then constructed from this reduced dataset, using Hilbert transforms to build the quadrature, as described in section \ref{algorithm}.
The resulting data is divided into equal segments of length $T$.
The computational cost of these operations is negligible, especially because they are only performed once at the beginning of the search.

Next, a bank of bandpass filtered segments is built.
We choose a filter half-bandwidth $B > \Delta f$, where $\Delta f$ is the maximum of the signal half-bandwidth over time $T$, and for every segment, we construct $N_f$ smaller filtered segments by using a heterodyne technique \cite{NRSSWD}.
These filtered segments are arranged to cover completely the frequency interval from $f_{\rm min}$ to $f_{\rm max}$.
If we impose the requirement that no signal power should be lost due to the filtering, the number $N_f$ is a function of $B-\Delta f$:
the whole frequency interval is divided into subintervals of length $2B$, displaced from each other by $2(B-\Delta f)$, so that any signal with half-bandwidth $\Delta f$ is guaranteed to be fully contained within one subinterval, and $N_f=(f_{\rm max}-f_{\rm min})/2(B-\Delta f)$.
In the heterodyne technique, the signal is multiplied by a complex exponential $\exp{2\pi i(B-f_0)}t$, low-pass filtered (for reasons of antialiasing), and resampled at frequency $8B$.
The resulting segments have $8BT$ (complex) data points, and their Nyquist frequency is $4B$.
The cost of the filtering, if implemented by an infinite impulse response digital filter, is of order $2(2n-1)$ floating point operation per input point, where $n$ is the number of poles in the filter, which is a function of the filter performance; we shall denote this cost by $C_{\rm f}$.
Resampling will have negligible computational cost compared to the multiplication by the complex exponential; assuming that the latter is not computed at run time, the cost for the construction of the filtered segments bank is
\begin{equation}
(C_{\rm f}+6) t_{\rm obs} \frac{(f_{\rm max}-f_{\rm min})^2}{2(B-\Delta f)} \: {\rm flop}. \label{eq:bcost}
\end{equation}
This cost can be adjusted by varying $B$ and $T$, which enters the formula through the dependence of $\Delta f$ on it.

Once the filtered segments bank is constructed, for each of the $N_p$ points in parameter space, and for every segment at time $t<t_{\rm obs}-T-{\rm d}/2-{\rm yr}/2$, we select the appropriate filtered segments from the bank, i.e. filtered segments where the signal is expected for times $t$, $t_1 = t+{\rm d}/2$, $t_2 = t+{\rm yr}/2$ and $t_3 = t+{\rm d}/2+{\rm yr}/2$, and we multiply them together.
At this point, the modulations induced from the detector motion have been removed; what remain to be applied are corrections for the spin-down of the source.
A refined gridding of the spin-down portion of parameter space, for the already quite narrow region used for the choice of filtered segments from the bank, is used to resample the product of the four segments according to the new time
\begin{equation}
t' = \frac{t f_0(t) + t_1 f_0(t_1) + t_2 f_0(t_2) + t_3 f_0(t_3)}{f_0(0)},
\end{equation}
where the origin of $t$ was arbitrarily set to zero.
The resampling will only affect a few points in the time series, so its cost is negligible in itself.
The power spectrum is constructed by taking the norm of the Fourier transform.
Finally, all the spectra corresponding to a given point in parameter space (from the partitioning of time in segments of length $T$) are added together to form the final spectrum, which is searched for significant peaks up to the Nyquist frequency $4B$.
All these operations are done on small segments of length $8BT$, and the total cost is the sum of the cost for building the products for all $N_p$ points and of the cost for the spin-down corrections, power spectra construction and sum for the $N_p'$ points corresponding to the finer grid:
\begin{equation}
B t_{\rm obs} N_p [72 + 20N_p'(1 + \log_2 8BT)] \: {\rm flop}. \label{eq:dcost}
\end{equation}
Again, this can be adjusted by varying $B$ and $T$, since $N_p$ and $N_p'$ depend on both $B$ and $T$.
The total cost of the search is the sum of equations \ref{eq:bcost} and \ref{eq:dcost}.

The number $N_p$ will be the number of non-intersecting (hyper-) volume elements in parameter space required to cover completely the region of interest, such that all points within a given volume element correspond to the same choice of filtered segments from the bank, at each of the four times of interest in this problem.
Manifestly, $N_p$ will depend on $B$ through the way it partitions the frequency interval $[f_{\rm min}, f_{\rm max}]$, and on $T$, which influences the size of a volume element over which the source frequency varies by some fixed amount.
We solve the problem of computing $N_p$ numerically, by building a mean volume element: we select at random a point in parameter space, and compute its associated volume element.
Repeating this procedure, we construct the average of a volume element, and divide the parameter space volume by it; the result is an estimate of $N_p$.

The refined gridding should only be necessary for searches with large spin-down rates (see below), and its effect will be to force to repeat $N_p'$ times the operations following and including the resampling above, where $N_p'$ is the number of points on the refined grid.
The computation of $N_p'$ is not of the same nature as the one for $N_p$: the problem now only involves the spin-down parameters, and the goal is to minimize the losses in signal power from residual frequency drifts, or misalignments of the spectra that are summed together.
In that sense, it is similar to the problem of computing the number of points in parameter space for the stack-slide technique \cite{BC}, and should be solved using the same geometrical approach.
However, we shall rather take $N_p'=1$, and justify this approximation by the relatively large spin-down time ($\tau=5000 {\rm yr}$) to be used in the numerical example of section \ref{numerical}.
It should be kept in mind that this may not be sufficient for faster spin-down, in the sense that under this approximation $N_p'$ may be under-estimated.
With the ability we now have to compute the required computational cost as a function of $B$ and $T$, the only thing that remains to be described is how we characterize the sensitivity of the search, as a function of the same choice of parameters.

\subsection{Sensitivity} \label{sensitivity}
From a statistical point of view, the results in \cite{Thorne,BCCS,BC} are presented in a somewhat non-standard way, so to ease comparison of our results with theirs, we first briefly reformulate their definition of sensitivity.
By setting the average noise power to be equal to one, we can express the signal amplitude $h$ in units of $\sqrt{S_n(f)/t_{\rm obs}}$, where $S_n(f)$ is the noise spectral density at frequency $f$.
In order to be consistent with \cite{Thorne,BCCS,BC}, we use the definition that 
the averaged signal power (also the square of the signal to noise ratio in amplitude) is $P_s=\langle F^2_+ \rangle h^2$, where $\langle F^2_+ \rangle=1/5$ is the average of the detector response over all angles and polarizations \cite{Thorne}.
This definition is useful to approximately account for variations in $h$ over time that are produced by the motion of the source in the detector response pattern.
For a true signal power $P_s$, computing the power spectrum of a time series as a sum of $n$ of its subseries spectra (this describes the statistics of both the coherent and stack-slide techniques) will give a distribution $p_n(P|P_s)$ of the power $P$ \cite{Groth}.
Given a false alarm probability $1-\alpha$, one defines a threshold power $P_\alpha$ such that the integral of $p_n(P|0)$ from 0 to $P_\alpha$ is $\alpha$.
Observations with power exceeding $P_\alpha$ will be called ``detections with confidence level $\alpha$'' \footnote{The authors of \cite{BCCS,BC} define the false alarm probability as $1-\alpha/N_p$, where $N_p$ is the number of points in parameter space.}.
One further defines a false dismissal probability $\beta$ as a function of the signal power, as the integral of $p_n(P|P_s)$ with respect to $P$, from 0 to $P_\alpha$.
It has been customary in the field \cite{Thorne,BCCS,BC} to use the relation $\langle P \rangle = P_s + n$ for the average observed power $\langle P \rangle$ to interpret $P_\alpha$ as a lower bound on the detectable signal amplitude, by setting $P_\alpha=\langle P \rangle$.
For instance, performances of sub-optimal techniques are frequently compared to the optimal case defined by $n=1$; for $\alpha=0.99$, the minimal detectable amplitude as defined above (the {\it optimal} amplitude) is $h=4.2$, and the corresponding false dismissal rate is $\beta \simeq 0.57$. 
In fact, it is easily checked that as $n$ gets larger, this definition gives a value of $\beta$ that approaches $1/2$ from above.
Therefore, for the ease of comparison with other works, we will define our minimal detectable amplitude as the smallest value of $h$ that corresponds to a false alarm probability $1-\alpha$ and a false dismissal rate $\beta=1/2$.
The sensitivity will be defined as $\Theta=4.2/h$, and this definition will be approximately compatible with those of \cite{Thorne,BCCS,BC}.

We use numerical analysis to determine $h$ as a function of $B$ and $T$.
We first set the signal amplitude to zero, and generate a large number of filtered segments of pure white noise, that are multiplied together and then added by groups of $(t_{\rm obs}-T-{\rm d}/2-{\rm yr}/2)/T$ (i.e., the number of segments in the portion of the observation period that can be used for multiplications).
By constructing the cumulative probability function for the resulting segments, we then evaluate the threshold $P_\alpha$.
We repeat the procedure for non-zero signal amplitudes: picking at random in a uniform distribution the parameters of the source, we construct the final spectrum, and evaluate the power in the expected frequency channel.
For each choice of signal amplitude, we repeat this a large number of times for different source parameters, and build a distribution of the power in the expected channels; the values of $\beta$ as a function of signal power are then deduced from these distributions and the previously calculated value of $P_\alpha$.
Therefore, the sensitivity $\Theta$ is effectively averaged over all possible source parameters.
To reduce computational burden, spin-down effects are not directly included, in the sense that we only consider them in determining the bandwidth of the signal for a given $T$, and do not include them in the initial simulated segments that are multiplied together.
By doing so, we might reduce the non-linear mixing of signal with noise, and therefore deduce a sensitivity that is too high; however, the magnitude of this mixing is such that the error so induced is expected to be small.

\section{Numerical Simulations} \label{numerical}
\begin{figure}
\begin{center}
\epsfbox{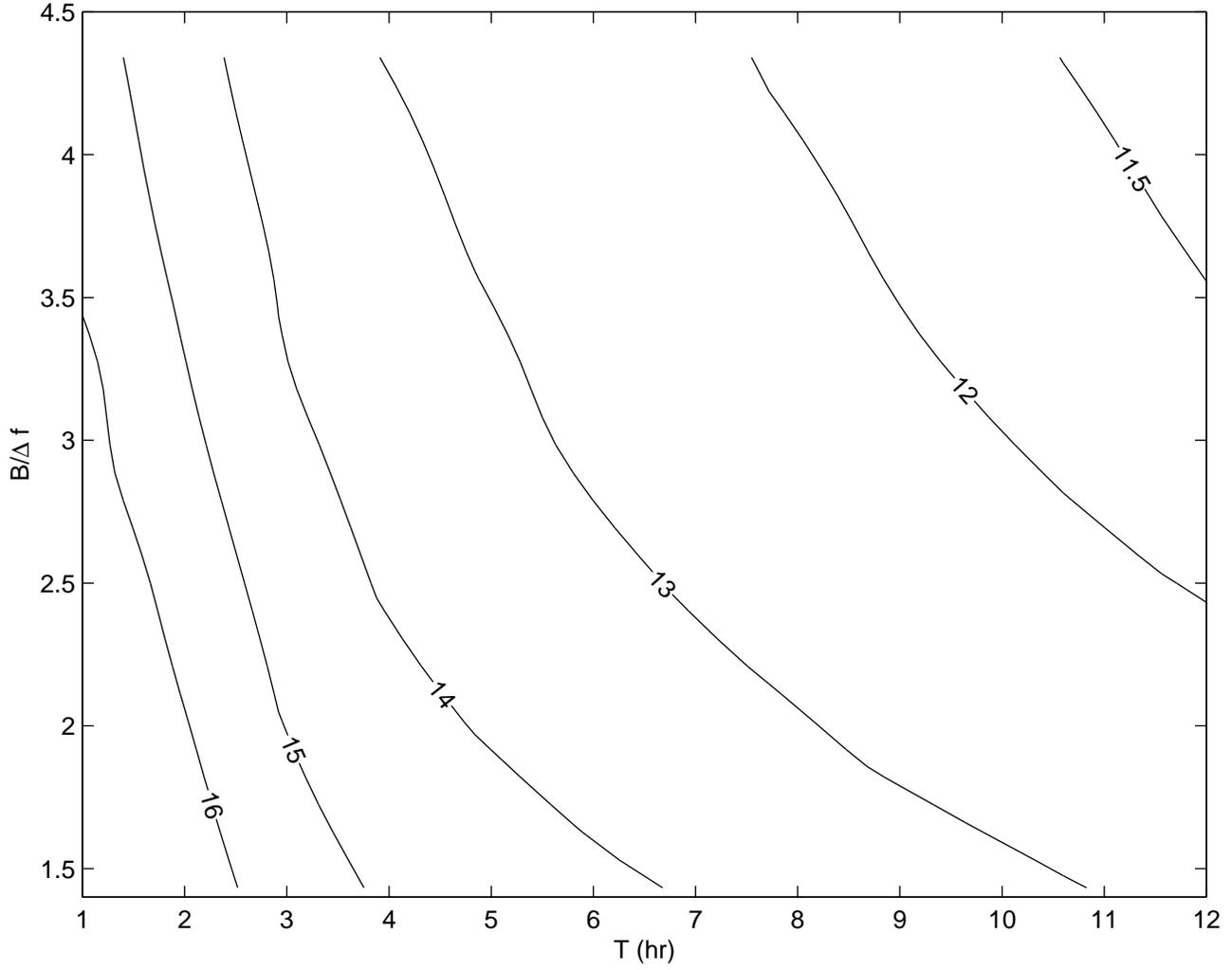}
\end{center}
\caption{A contour plot of $\log_{10} N_p$ as a function of $B$ (in units of eq.~\ref{eq:signal_bandwidth}) and $T$. Labels on the contour lines are values of $\log_{10} N_p$.}
\label{fig:Np}
\end{figure}

\begin{figure}
\begin{center}
\epsfbox{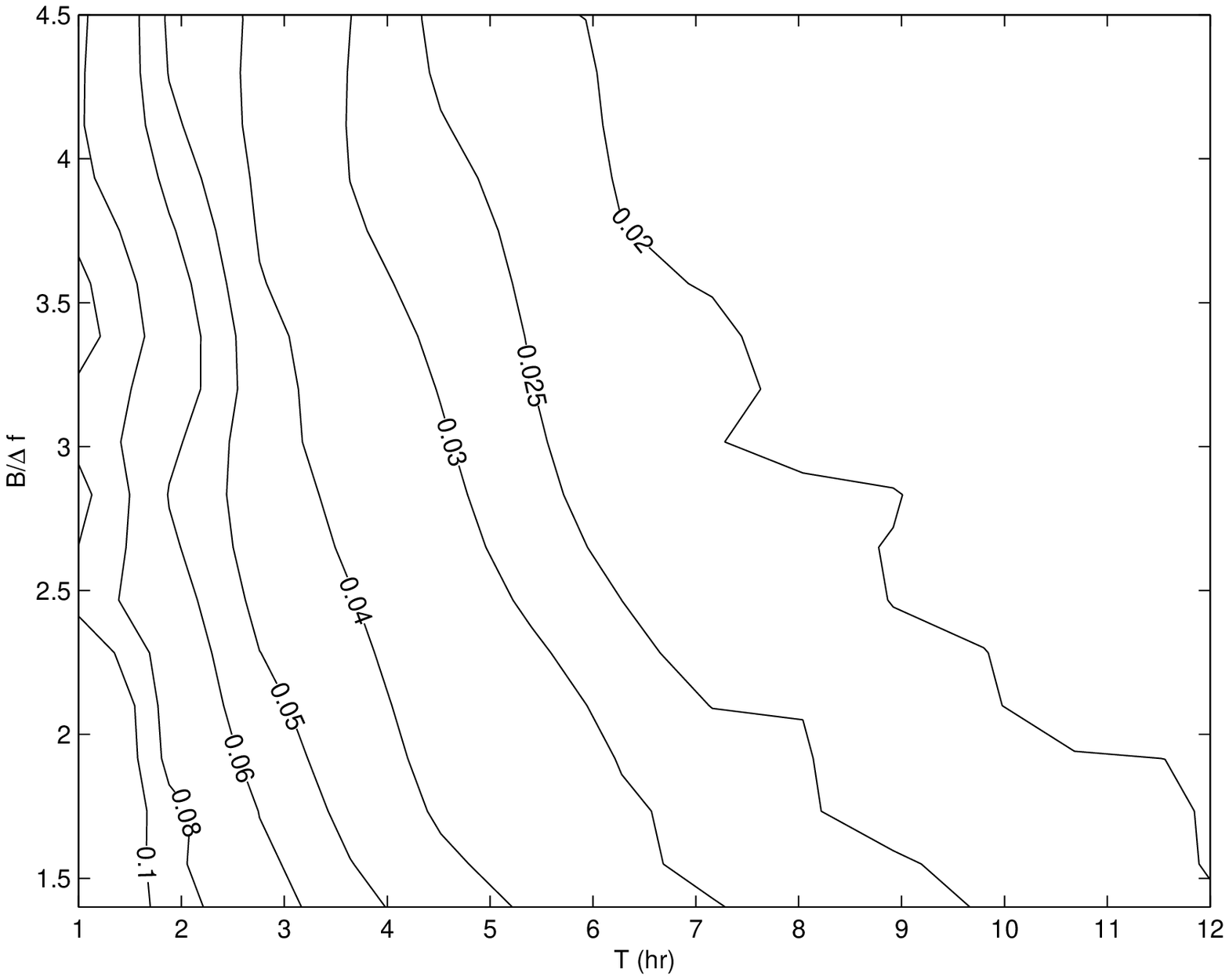}
\end{center}
\caption{A contour plot of $\Theta$ as a function of $B$ (in units of eq.~\ref{eq:signal_bandwidth}) and $T$. Labels on the contour lines are values of $\Theta$.}  
\label{fig:h}
\end{figure}

We opted for a relatively ``easy'' {\it broad}-band all-sky search for our explorations. 
We choose $f_{\rm min} = 40 {\rm Hz}$, $f_{\rm max} = 1 {\rm kHz}$, $\tau=5000 {\rm yr}$, $t_{\rm obs} = 1 {\rm yr}$, and confidence level $\alpha=99\%$.
Note that $t_{\rm obs} = 1 {\rm yr}$ is the minimal possible observation time, and that the data at times $t > {\rm d}/2+{\rm yr}/2$ can not be demodulated.
This obviously limits the sensitivity of the search, although it does so in a diminushing proportion as $t_{\rm obs}$ grows above one year.
A different {\it narrow}-band all-sky search was also considered, with parameters as for the broad-band search, except for $f_{\rm min} = 450 {\rm Hz}$ and $f_{\rm max} = 500 {\rm Hz}$, perhaps corresponding to the case of an interferometer made narrow-band using signal recycling. 
For definitiveness, we choose the latitude of the detector to correspond to the LIGO Hanford detector.
As described above, we computed numerically the value of $N_p$ for different values of $T$ and $B$ (from now on, $B$ will be expressed in units of $\Delta f(T)$ as defined in eq.~\ref{eq:signal_bandwidth}, with $f_0 = f_{\rm max}$, and $A_1$, $A_2$ averaged over all angles $\delta$); see figure \ref{fig:Np}.
Using the method described in section \ref{sensitivity}, we computed the dependence of the relative sensitivity, $\Theta$, on the same variables; the results are shown in figure \ref{fig:h}.

\begin{figure}
\begin{center}
\epsfbox{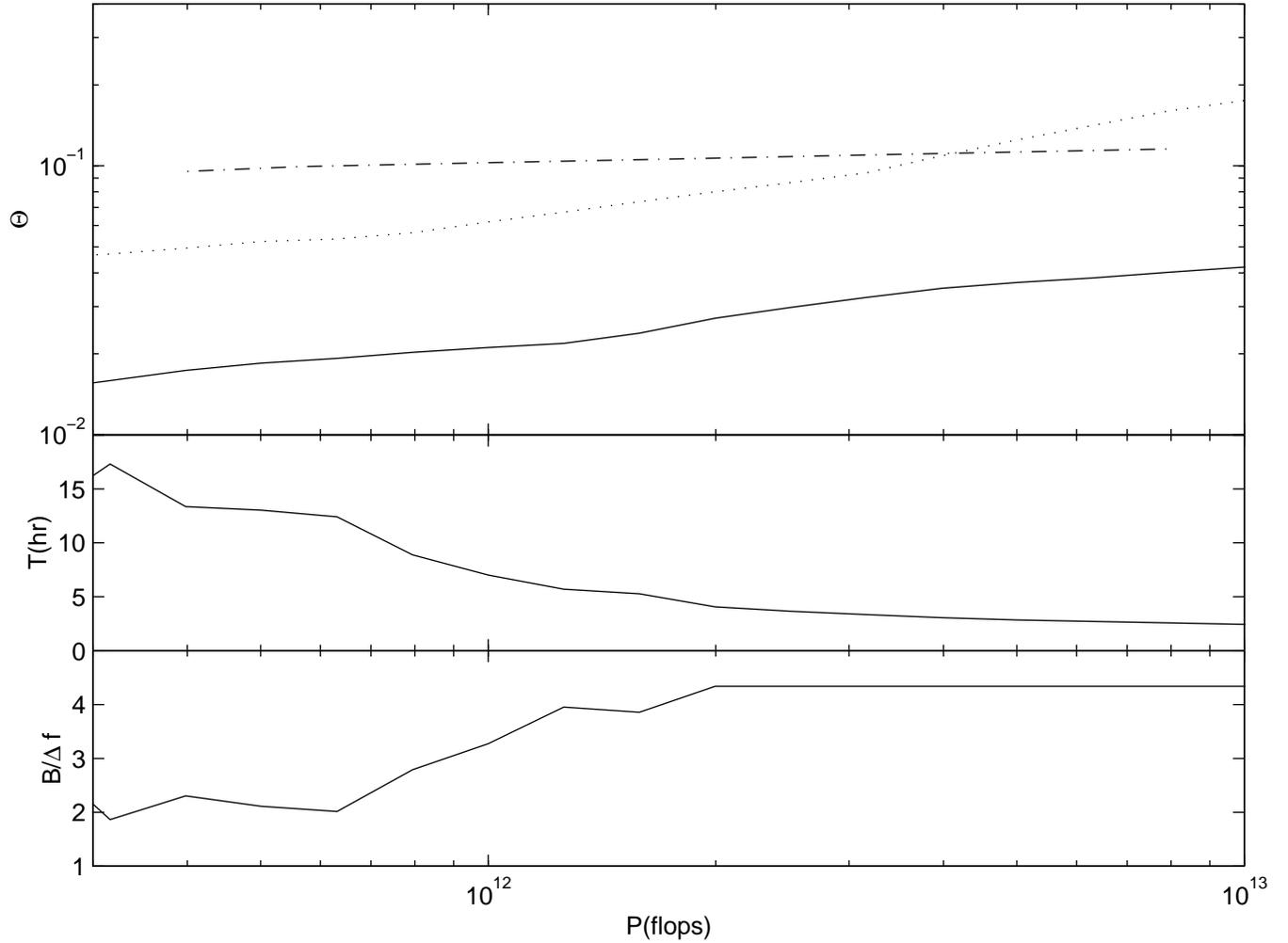}
\end{center}
\caption{Top: the relative sensitivity $\Theta$ for the optimized algorithm, as a function of computational power $P$ (solid line), and the corresponding for the stack-slide technique (dash-dotted line), both for the broad-band search.
The dotted line corresponds to the different narrow-band search. 
Middle: the optimal segment length $T$ in the broad-band search. 
Bottom: the optimal filter bandwidth $B$ (in units of eq.~\ref{eq:signal_bandwidth}), also in the broad-band search.}  
\label{fig:Theta}
\end{figure}

To get a meaningful notion of the computational power $P$ from the cost of the search (the sum of equations \ref{eq:bcost} and \ref{eq:dcost}), we imposed the requirement that the analysis should be done in a time $t_{\rm obs}$. 
Our map of $N_p$ vs.\ $B$ and $T$ was then reexpressed as a map of $P$ versus $B$ and $T$.
We observed that for plausible filter implementations, the contribution to $P$ from eq.\ \ref{eq:dcost} was at least two orders of magnitude larger than the contribution from eq.\ \ref{eq:bcost} in the region of interest of the $T,B$--plane; our results are highly insensitive to the real cost of constructing the filtered segments bank.
Finally, we optimized our algorithm for a given computational power by choosing the values of $B$ and $T$ that maximize $\Theta$ along a line of constant $P$.
The results are shown in figure \ref{fig:Theta}, together with the corresponding sensitivity for the stack-slide technique.

\subsection{Discussion}
Fig.\ \ref{fig:Theta} shows that the sensitivity of the algorithm described in this paper is roughly 30\% of the sensitivity of the stack-slide technique in a broad-band search, for computational power ranging from 0.4 Tflops to 8 Tflops.
For the same range of computational power, the minimum amplitude detectable with 50\% efficiency ranges approximately from 25 to 60 times the optimal amplitude.

Also shown in fig.\ \ref{fig:Theta} is the sensitivity of an easier narrow-band all-sky search. 
When the computational power reaches 4 Tflops, the sensitivity eventually gets better than that of the stack-slide technique, as computed for the harder, broad-band search.
However, it is not clear how the performances of the stack-slide technique scale with the bandwidth of the search, although they do scale strongly with the upper frequency.
It is not possible, for instance, to apply the stack-slide technique on a heterodyned dataset without having to search over the additional source frequency parameter, because of the coupling of the source frequency to both the detector rest frame time and to the phases introduced by the detector motion ($M_1$ and $M_2$ in equation \ref{eq:phase_fnc}).
The comparison of our algorithm to the stack-slide technique is therefore not trivial, and in particular there might be some narrow-band searches where the stack-slide performances are approached or outperformed.

Our results suggest that the distortions in the noise produced by the nonlinearities inherent to our algorithm can not be alleviated by narrow-banding the signal such that the algorithm performs in a manner that is satisfactory for a realistic broad-band all-sky search.
The need for narrow-banding adds one dimension (the source frequency) to the parameter space, and this greatly increases the computational burden.
Moreover, the dependence of the sensitivity on the bandwidth of the bandpass filter, which is the most important parameter in determining the number of points in parameter space, is strong enough that efficient detection is computationally intensive.
Consequently, our principal conclusion is that, although the symmetry described in section \ref{algorithm} simplifies the problem from a formal point of view, it does so by strongly increasing the level of noise, principally by mixing components of noise of different frequencies.
We have presented an algorithm that minimizes this effect, but have found that for a realistic broad-band search, it did not perform better than the stack-slide technique at the same computational power.

Most detection algorithms for periodic sources, and in particular the stack-slide technique, probably don't have a strong scaling of their performances with the bandwidth of the search, essentialy because making use of the reduced bandwidth adds complexity by forcing one to consider the frequency as an explicit parameter of the search.
It is therefore possible that our algorithm compares more advantageously to these other algorithms in narrow-band searches, which can be considered as more natural problems for it, because it necessarily involves a search over the source frequency.

\acknowledgments The author wishes to thank Rainer Weiss for helpful discussions and comments on the work presented here. This work was supported by the National Science Foundation under cooperative agreement PHY-9210038, and by the {\it Fonds pour la Formation de Chercheurs et l'Aide \`a la Recherche} of the Province of Qu\'ebec, Canada.


\begin{references}
\bibitem{LIGO_VIRGO} A. Abramovici {\it et al.}, Science {\bf 256}, 325 (1992); C. Bradachia {\it et al.}, Nucl. Instrum. Methods Phys. Res. A {\bf 289}, 518 (1990).  

\bibitem{Thorne} K. S. Thorne, in {\it 300 Years of Gravitation}, edited by S. W. Hawking and W. Israel (Cambridge University Press, Cambridge, England, 1987), pp. 330-458.

\bibitem{DV} Dhurandhar, S. V., Vecchio, A., gr-qc/9911102 (1999); Dhurandhar, S. V., Vecchio, A., gr-qc/0011085 (2000).

\bibitem{JKS} P. Jaranowski, A. Kr\'olak, B. F. Schutz, Phys. Rev. D {\bf 58}, 063001 (1998); P. Jaranowski and A. Kr\'olak, Phys. Rev. D {\bf 59}, 063003 (1999); P. Jaranowski and A. Kr\'olak, gr-qc/9901013 (1999).

\bibitem{BCCS} P. R. Brady, T. Creighton, C. Cutler, B. F. Schutz, Phys. Rev. D {\bf 55}, 2101 (1998).

\bibitem{BC} P. R. Brady, T. Creighton, Phys. Rev. D {\bf 61}, 082001 (2000).

\bibitem{SP} B. F. Schutz, M. A. Papa, gr-qc/9905018 (1999).

\bibitem{Hahn} S. L. Hahn, {\it Hilbert transforms in signal processing} (Artech House Ed., Boston, USA, 1996). 

\bibitem{Groth} E. J. Groth, Astrophys. J. Supp. Ser. No. 286, {\bf 29}:285-302 (1975).

\bibitem{NRSSWD} For examples of such a heterodyne technique, see: T. M. Niebauer {\it et al.}, Phys. Rev. D {\bf 47}, 3106 (1991).

\end{references}
\end{document}